\documentclass[showpacs,preprintnumbers,amsmath,amssymb,prd,twocolumn,nofootinbib]{revtex4}

\usepackage{graphicx}% Include figure files
\usepackage{dcolumn}% Align table columns on decimal point
\usepackage{bm}% bold math
\def\comment#1{}

\def\slashchar#1{\setbox0=\hbox{$#1$}           % set a box for #1 
   \dimen0=\wd0                                 % and get its size
   \setbox1=\hbox{/} \dimen1=\wd1               % get size of /
   \ifdim\dimen0>\dimen1                        % #1 is bigger
      \rlap{\hbox to \dimen0{\hfil/\hfil}}      % so center / in box
      #1                                        % and print #1
   \else                                        % / is bigger
      \rlap{\hbox to \dimen1{\hfil$#1$\hfil}}   % so center #1
      /                                         % and print /
   \fi}                                         %

\def\sigmab{{\mbox{\boldmath $\sigma$}}}

\begin{document}

\title{Universal properties of the $U(1)$ current at deconfined quantum critical points: comparison with 
predictions from gauge-gravity duality}

\author{Flavio S. Nogueira}
\email{nogueira@physik.fu-berlin.de}
\affiliation{Institut f{\"u}r Theoretische Physik,
Freie Universit{\"a}t Berlin, Arnimallee 14, D-14195 Berlin, Germany}

\date{Received \today}

\begin{abstract}
The deconfined quantum critical point of a two-dimensional $SU(N)$ antiferromagnet is governed by an Abelian Higgs model 
in $d=2+1$ spacetime dimensions featuring $N$ 
complex scalar fields. In this context, we derive for $2\leq d\leq 4$ 
an exact formula for the central charge of the $U(1)$ current in terms of the gauge coupling at quantum criticality and 
compare it with the corresponding result obtained using gauge-gravity duality. 
There is a remarkable similarity precisely for $d=2+1$. In this case the amplitude of the current correlation function has 
the same form as predicted by the gauge-gravity duality. We also compare finite temperature results for 
the charge susceptibility in the large $N$ limit with the result predicted by the gauge-gravity duality. 
Our results suggest that condensed matter systems at quantum criticality 
may provide interesting quantitative tests of the gauge-gravity duality even in absence of supersymmetry.
\end{abstract}

\pacs{64.70.Tg, 11.10.Kk, 11.25.Tq}
\maketitle

Correlation functions involving conserved physical quantities are in some special circumstances  
determined by simple dimensional analysis. This is the case for a theory at its quantum critical point. 
For relativistically invariant theories having a conserved $U(1)$ current $j_\mu(x)$, the current correlation function  
at criticality has a remarkable simple form, being given by \cite{Note-1}
\begin{equation}
\label{cc}
 \langle j_\mu(x)j_\nu(0)\rangle=\frac{k}{S_d^2x^{2(d-1)}}\left(\delta_{\mu\nu}-\frac{2x_\mu x_\nu}{x^2}\right),
\end{equation}
where $S_d=2\pi^{d/2}/\Gamma(d/2)$ is the surface of a unit sphere in $d$ spacetime dimensions. The above result follows from 
the conformal invariance of the quantum critical point and the fact that (on-shell) conserved $U(1)$ currents do not have an anomalous dimension. 
Thus, the dimensionless number $k$ is the central charge of the $U(1)$ current in a $d$-dimensional conformal field theory (CFT$_d$). 

Another dimensionless number, $k'$, is defined at a quantum critical point and finite temperature via the charge susceptibility 
$\chi=\lim_{V\to\infty}\langle Q^2\rangle/V$, where $Q$ is the conserved charge associated to the $U(1)$ current, i.e., it is simply 
obtained by integrating $j_0(x)$ over $d-1$ spatial dimensions. Thus, dimensional analysis and universality implies
\begin{equation}
 \chi=k'T^{d-2}.
\end{equation}

Recently, Kovtun and Ritz \cite{Kovtun-Ritz} have shown that for theories having a gravitational dual \cite{AdS/CFT,AdS/CFT-PT}, the 
following relation holds, 
\begin{equation}
\label{ratio-KR}
 r_G(d)\equiv\frac{k'}{k}=\frac{1}{2\pi^{d/2}}\left(\frac{4\pi}{d}\right)^{d-2}\frac{\Gamma^3(d/2)}{\Gamma(d)},
\end{equation}
which for $d=2$ agrees with the exact result for a CFT$_2$ \cite{Kovtun-Ritz}, namely,
\begin{equation}
\label{ratio-KR-2d}
 \left.\frac{k'}{k}\right|_{d=2}=\frac{1}{2\pi},
\end{equation}
the latter being complementary to a well-known relation for the central charge of the energy-momentum tensor, $c$, and the universal amplitude of the free-energy 
\cite{Affleck-1986,Cardy-1986}. 

It would be interesting to find examples where the result (\ref{ratio-KR}) holds. 
Note that this would not necessarily imply that a theory fulfilling (\ref{ratio-KR}) has a gravitational dual, but finding such an 
example would be very encouraging. 
In this paper we will show that condensed matter systems at quantum criticality may provide relatively simple quantitative  
tests of gravitational duality. We will derive results for $k$ and $k'$ in effective theories 
of quantum antiferromagnets. These results will then be compared with those predicted by the gauge-gravity duality. 

The quantum critical point of an $SU(N)$ quantum antiferromagnet is described by an Abelian Higgs model with $N$ complex bosons 
\cite{Sachdev-Review,Senthil-2004}, i.e., 
\begin{eqnarray}
\label{L-boson}
{\cal L}&=&\frac{1}{4}F_{\mu\nu}^2+
\sum_{a=1}^{N}|D_\mu z_\alpha|^2\nonumber\\
&+&r_0\sum_{a=1}^{N}|z_\alpha|^2+\frac{u_0}{2}
\left(\sum_{a=1}^{N}|z_\alpha|^2\right)^2,
\end{eqnarray}
where $D_\mu=\partial_\mu+ie_0A_\mu$. 
Obviously, the conserved $U(1)$ current for the above Lagrangian is 
\begin{equation}
 j_\mu=-ie_0\sum_{a=1}^N(z_a^*D_\mu z_a-z_aD_\mu^* z_a^*).
\end{equation}
For a spin 1/2 antiferromagnet, the Lagrangian (\ref{L-boson}) follows from the representation of the spin orientation field as ${\bf n}(x)=z_a^*(x)\sigmab_{ab}z_b(x)$, 
where the fields $z_a$ represent the so called spinons and 
$\sigmab=(\sigma_1,\sigma_2,\sigma_3)$ is a vector having the Pauli matrices as components. The generalization to $SU(N)$ follows simply by replacing the 
Pauli matrices by the generators of $SU(N)$. 
From this representation a local gauge symmetry naturally 
emerges, since the spin orientation field ${\bf n}(x)$ remains invariant under the local phase transformation $z_a(x)\to e^{i\theta(x)}z_a(x)$. 
Perturbatively, this theory has a quantum critical point for large enough $N$ and  
$2<d<4$ \cite{Nogueira_2007,Nogueira_2008}, in which case the spinons $z_a$ are deconfined in the sense 
defined precisely in Ref. \cite{Senthil-2004}. A quantum critical 
point has been found non-perturbatively for all $N$ by means of the exact RG \cite{Litim}. 
For $d=3$ spacetime dimensions this quantum critical point separates two distinct Mott insulating phases. In one of these phases the 
fields $z_a$ are condensed, leading in this way to  
antiferromagnetic (AF) order. This corresponds to a Higgs phase in particle physics terminology. The other phase is paramagnetic and corresponds  
in the lattice to an insulating pattern formed by a crystalline structure of singlet bonds, the so called valence-bond solid (VBS) \cite{RS}. 
In particle physics language, this phase corresponds to a spinon confinement phase. 
The character of the AF-VBS phase transition for $SU(2)$ spins is at present controversial, with some numerical results in the lattice favoring a 
second-order phase transition (and therefore, the existence of a quantum critical point) \cite{JQ-Model}, while other numerical results favor a weak first-order 
phase transition \cite{Kuklov_2008}.
In this paper we will assume that a quantum critical point exists, which is definitly true if $N$ is large enough and $2<d<4$ \cite{Note-3}. 
Precisely at the quantum critical point the theory is a CFT$_d$. 

Let us show that there exists an exact relation between the number $k$ in Eq. (\ref{cc}) and the dimensionless 
renormalized gauge coupling at the quantum critical point. 
The renormalized gauge coupling is uniquely determined by the vacuum polarization, as dictated by the Ward identities. 
The vacuum polarization, on the other hand, is related in momentum space to the current correlation function by
\begin{equation}
 \Pi(p)=-\frac{e_0^2}{(d-1)p^2}\langle j(p)\cdot j(-p)\rangle,
\end{equation}
where $e_0^2$ is the bare gauge coupling. Therefore, the renormalized gauge coupling $e^2(p)$ is given by the 
exact expression
\begin{equation}
\label{e2} 
\frac{1}{e^2(p)}=\frac{1}{e^2_0}-\frac{1}{(d-1)p^2}\langle j(p)\cdot j(-p)\rangle.
\end{equation}
We are interested in the regime near or at the quantum critical point, so that $|p|\ll e_0^2$, which is the regime where 
Eq. (\ref{cc}) is valid. 
From Eq. (\ref{cc}) it is easy to obtain that
\begin{equation}
\label{cc-p}
 \langle j(p)\cdot j(-p)\rangle=\frac{(d-2)\Gamma(1-d/2)\Gamma^2(d/2)}{(4\pi)^{d/2}\Gamma(d-1)}~k|p|^{d-2}.
\end{equation}
The dimensionless gauge coupling is defined by $f(p)=p^{d-4}e^2(p)$. 
Its critical value is given by the RG fixed point $f_*=\lim_{p\to 0}f(p)$. 
Therefore, we obtain from Eqs. (\ref{e2}) and (\ref{cc-p}) that $k$ 
is {\it exactly} related to the fixed point $f_*$ via the formula: 
\begin{equation}
\label{f*}
 k=\frac{2^{d-1}\pi^{d/2}\Gamma(d)}{\Gamma(2-d/2)\Gamma^2(d/2)}\frac{1}{f_*}.
\end{equation}

A similar relation holds in   
theories admitting a gravitational dual, with the difference that the central charge of the theory in the boundary is related to 
the gauge coupling of the theory in the bulk, which is also coupled to gravity in an AdS background. 
Indeed, an expression for the central charge $k$ in terms of the gauge coupling of the   
Einstein-Maxwell action in a $(d+1)$-dimensional anti-de Sitter (AdS$_d$) spacetime was derived by Freedman {\it et al.} \cite{Freedman_1999}. 
The Einstein-Maxwell action is 
\begin{eqnarray}
\label{S-EM}
 S&=&\frac{1}{16\pi G}\int d^{d+1}x\sqrt{-g}[R+d(d-1)\Lambda]\nonumber\\
&-&\frac{1}{4g_{d+1}^2}\int d^{d+1}x\sqrt{-g}F_{\mu\nu}^2.
\end{eqnarray}
The cosmological constant in the above action sets the scale to define the dimensionless gauge coupling 
as $\hat g_{d+1}^2=\Lambda^{(d-3)/2}g_{d+1}^2$. The $U(1)$ central charge $k$ of the CFT$_d$ on the boundary is related to the gauge coupling 
of the theory in the bulk by \cite{Freedman_1999}
\begin{equation}
\label{k-AdS}
 k=\frac{2\pi^{d/2}(d-2)\Gamma(d)}{\Gamma^3(d/2)}\frac{1}{\hat g_{d+1}^2}.
\end{equation}
Note that, in contrast with Eq. (\ref{f*}), where $f_*$ is the coupling constant of the CFT$_d$, in Eq. (\ref{k-AdS}) 
$\hat g_{d+1}^2$ is the coupling constant of the theory with gravity.  
Remarkably, Eqs. (\ref{f*}) and (\ref{k-AdS}) have exactly the same form for $d=3$ 
spacetime dimensions, where $g_4^2$ is dimensionless:
\begin{equation}
\label{comparison-ks}
 k=\frac{32}{f_*},~~~~~~~~~k=\frac{32}{g_{4}^2}.
\end{equation}
Despite the similarity between the two $k$'s, we cannot safely claim that $f_*$ can be identified with $g_4^2$, since 
unfortunately the 
symmetries of the quantum antiferromagnet do not entirely match with the ones of the supergravity theory. 
Another important issue is that usually in applications involving the quark-gluon plasma \cite{Son-Starinets-Review} or quantum critical phenomena in 
condensed matter physics \cite{Herzog-2007,Hartnoll,Hartnoll-Review} the current is associated to a global symmetry of the boundary theory. Here, in order to relate 
$k$ to the fixed point $f_*$, we have considered a current having a local gauge symmetry. In $SU(N)$ supersymmetric QCD, where the 
$\beta$ function is exactly known \cite{NSVZ}, many other interesting results for the central charges 
of gauge invariant currents are known \cite{Anselmi-NPB-1998}.  

As a concrete example, let us consider the large $N$ limit of the theory (\ref{L-boson}).  
In this case we have in $d$ spacetime dimensions, 
\begin{equation}
\label{FP}
 f_*=\frac{(4\pi)^{d/2}(d-2)\Gamma(d)}{4\Gamma(2-d/2)\Gamma^2(d/2)N},
\end{equation}
which is just the fixed point of the RG one-loop $\beta$ function for a fixed dimensionality \cite{Kleinert-Nogueira-2}.  
Thus, Eq. (\ref{f*}) yields  
\begin{equation}
\label{k-free}
 k=\frac{2N}{d-2},
\end{equation}
which is the same result as for a free scalar field theory featuring $N$ complex fields. 
Up to the factor $N$,  
this is also the same result as in the conformally invariant $O(n)$ model \cite{Petkou-1996,Note-2}. 
For $d=3$ the fixed point (\ref{FP}) becomes $f_*=16/N$. The four-dimensional gauge coupling constant of 
the gravitational theory, on the other hand, is given by 
$g_4^2=6\pi/(\sqrt{2}N^{3/2})$ \cite{Herzog-2007}. Thus, we see that the gravitational theory yields a non-perturbative value of 
$k$, since in this case $k=16\sqrt{2}N^{3/2}/(3\pi)$ instead of giving a result $\sim{\cal O}(N)$. 

If in addition we minimally couple the theory in (\ref{L-boson}) to $N_f$ Dirac fermions (with four-component spinors), corresponding to the algebraic charge liquid 
\cite{Kaul,Nogueira_2008,Kaul-Sachdev,Note-3}, 
we obtain for large $N$ and $N_f$, with $N/N_f$ arbitrary, the result
\begin{equation}
k=\frac{2N+4(d-2)N_f}{d-2}. 
\end{equation}
Thus, for the purely fermionic case we have $k=4N_f$.  

The above results can also be applied to the easy-plane system, by generalizing the global $U(1)\times U(1)$ symmetry 
to $O(N)\times O(N)$ with $N$ even \cite{Nogueira_2007}. Although the easy-plane antiferromagnet exhibits 
a first-order phase transition \cite{Kuklov_2006} (this is true for all $N$; see Ref. \cite{Nogueira_2007}), the central charge 
$k$ depends only on the fixed point $f_*$ and not on the other couplings of the theory. In this case a real value of $f_*$ can always 
be found for large enough $N$ in a minimal subtraction scheme. 
Indeed, the $\beta$ function of the gauge coupling $f$ is 
only a function of $f$, not depending on the other couplings of the theory. 
Resummation schemes actually indicate that $f_*$ may exist for all values of $N$ \cite{Folk}. 

Let us give a clear example of a situation in gauge theories 
where a universal constant appears even in the absence of a second-order phase transition. A simple example 
is provided by the static particle-antiparticle potential calculated from the Wilson loop.  
Quite generally, the static interquark potential is given 
at large distances by
%
%\begin{equation}
%\label{Potential}
 $V(L)=\tau_s L-C/L+{\cal O}(1/L^2)$,
%\end{equation}
where $\tau_s$ is the string tension and $C$ is a universal constant. It turns out that the constant $C$ is 
always universal, even if no quantum 
critical points exist, like for example in the case of Polyakov's compact Maxwell electrodynamics \cite{Polyakov}, where 
no phase transition happens, i.e., the string tension is always nonzero. The reason for this 
can be stated in very simple terms \cite{Peskin}: the contribution $\propto T/L$ in the large $L$ expansion of 
$TV(L)$ is invariant under a scale transformation $T\to\lambda T$, $L\to\lambda L$. For a bosonic string model of quark confinement we have, 
for example, $C=(d-2)\pi/24$ \cite{Luescher}. The constant $C$ is related to the current correlation function. Indeed, a simple 
calculation for an Abelian gauge theory yields
\begin{equation}
 C=\frac{S_d(d-2)f_*}{2^d\pi^{d-1}}.
\end{equation}
Although in the seld-dual easy-plane case, no quantum critical point arises, it is possible to find a zero for the $\beta$ function of the 
gauge coupling, so that $C$ will be a universal number. Similarly, other relations involving the current correlation function hold.  
For instance, the universal relations derived in Ref. \cite{Herzog-2007} for the self-dual easy-plane antiferromagnet seem to remain valid, although the  
system, at zero temperature, exhibits a first-order phase transition \cite{Kuklov_2006,Nogueira_2007}. 

An exact computation of $k'$ in terms of $f_*$ is not as straightforward as in the case of the central charge $k$ of the $U(1)$ current. 
However, since at large $N$ the central charge $k$ has the same value as in a free theory 
with $N$ complex scalar fields 
[an $O(2N)$-invariant theory], it is 
instructive to calculate $k'$ for this  
free theory and compute the ratio $k'/k$. The calculation of $\chi$ is more easily done by coupling $j_0(x)$ to 
a uniform source $h$ and computing $\chi=-\partial^2 f/\partial h^2|_{h=0}$, where $f$ is the free-energy density at the presence of the 
external source. This leads us to the result
\begin{equation}
 \chi=4NT\sum_{n=-\infty}^\infty\int\frac{d^{d-1}p}{(2\pi)^{d-1}}\frac{\omega_n^2}{(\omega_n^2+{\bf p}^2)^2},
\end{equation}
where $\omega_n=2\pi nT$. By performing explicitly the Matsubara sum and the momentum integral, we obtain
\begin{equation}
\label{k1-free}
 k'=N\pi^{(d-5)/2}(d-3)\Gamma\left(\frac{3-d}{2}\right)\zeta(3-d).
\end{equation}
In particular, for $d=3$ we have $k'=N/\pi$. 
Therefore, we obtain the ratio between the $k'$ above and the $k$ given in Eq. (\ref{k-free}) as
\begin{equation}
 r_{\rm free}(d)\equiv\frac{k'}{k}=\frac{\pi^{(d-5)/2}(d-2)(d-3)}{2}\Gamma\left(\frac{3-d}{2}\right)\zeta(3-d).
\end{equation}
This is the {\it exact} result for a free theory. Note that for $d=2$ it agrees with Eq. (\ref{ratio-KR-2d}), as it should. Interestingly, 
the $d=2$ result is in this case the same as the $d=3$ one, i.e., $r_{\rm free}(2)=r_{\rm free}(3)=1/(2\pi)$. This actually reflects a more general 
property of $r_{\rm free}(d)$, namely, 
\begin{equation}
 r_{\rm free}(d)=r_{\rm free}(5-d).
\end{equation}

The prediction for $k'$ from the AdS$_{d+1}$/CFT$_d$ correspondence is \cite{Kovtun-Ritz}
\begin{equation}
\label{kp-AdS}
 k'=\frac{d-2}{\hat g_{d+1}^2}\left(\frac{4\pi}{d}\right)^{d-2},
\end{equation}
which when divided by Eq. (\ref{k-AdS}) gives the relation (\ref{ratio-KR}). Note that $r_G(3)=\pi/24<r_{\rm free}(3)$. 
In Fig. 1 we plot the ratios $r_G(d)$ and $r_{\rm free}(d)$ in the range $2\leq d\leq 4$. We see from the figure that 
$r_G(d)\leq r_{\rm free}(d)$ in the interval $2\leq d\leq 4$. 

\begin{figure}
\includegraphics[width=8cm]{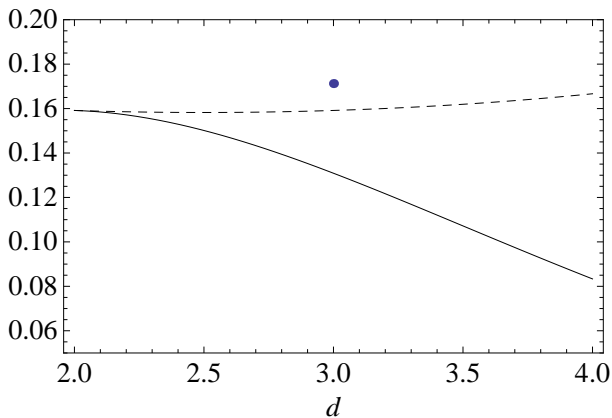}
\caption{Comparison between the ratios $r_G(d)$ and $r_{\rm free}(d)$. The continuous line represents 
$r_G(d)$, while the dashed line corresponds to $r_{\rm free}(d)$. The point represents the ratio $k'/k$ for 
the $O(n)$ non-linear $\sigma$-model at large $n$ and $d=3$ [see Eq. (\ref{r-nlsm})].}
\end{figure} 

The value of the ratio $k'/k$ for the $O(n)$ non-linear $\sigma$-model \cite{Note-2} at large $n$ and $d=3$ was 
given in Ref. \cite{Kovtun-Ritz} as \cite{CSY}
\begin{equation}
\label{r-nlsm}
 r_{O(n)}=\frac{\sqrt{5}}{2\pi}\ln\left(\frac{\sqrt{5}+1}{2}\right),
\end{equation}
which is about 7\% larger than $r_{\rm free}(3)$. Thus, for $d=3$ the free theory result is closer to 
the gravitational dual prediction than the corresponding large $n$ result for the $O(n)$ non-linear $\sigma$-model. 

In Ref. \cite{Klebanov-Polyakov} (see also Ref. \cite{Petkou-2003}) it was conjectured that there is a duality 
between the critical interacting $O(n)$ model 
in $d=3$ and an even spin gauge theory defined in AdS$_4$. It would be interesting to see 
such a gravitational dual construction working also in the case of the gauge theory formulation of quantum antiferromagnets. 

In summary, we have seen that at zero temperature the quantum critical gauge coupling of an $SU(N)$ quantum 
antiferromagnet is related via an exact formula to the central charge $k$ of the conserved $U(1)$ current 
[Eq. (\ref{f*})]. There is a 
corresponding formula for theories having a gravitational dual \cite{Freedman_1999} [Eq. (\ref{k-AdS})]. However, it relates the 
gauge coupling of a $(d+1)$-dimensional gravitational theory to the central charge of the $U(1)$ current in a  
confomal theory living at the boundary. These expressions have the same form for $d=3$, where the gauge 
coupling of the gravitational theory is dimensionless. At finite temperature we have computed the amplitude $k'$ of the 
charge susceptibility at quantum criticality in the large $N$ limit, which in this case is the same as 
the exact result for a model consisting of $N$ complex scalar fields. 
The ratio $k'/k$ at large $N$ was compared with the 
prediction of the gauge-gravity duality. They both agree at $d=2$ with the exact result $k'/k=1/(2\pi)$ for two-dimensional conformal field theories 
\cite{Kovtun-Ritz}. On the other hand, the derived large $N$ result for $k'/k$ is found to differ at $d=3$ of about 18\% from the gauge-gravity dual prediction. It 
would be interesting to see if better approximations can get closer to the gravitational duality result. One promising approach in this 
case is an expansion in $\epsilon=4-d$, where one would be able to give results for finite $N$. Another future project is to derive an exact formula 
for $k'$ in terms of $f_*$, as we did with $k$. It would be interesting if such a formula for $d=3$ has a form similar to Eq. (\ref{kp-AdS}), in the same way 
as with $k$ in Eq. (\ref{comparison-ks}). However, it should be mentioned here that 
such a computation may lead instead to a breakdown of the universaly of the ratio $k'/k$. For instance, in the context of the AdS$_{d+1}$/CFT$_d$ 
correspondence, Ritz and Ward \cite{Ritz-Ward} have recently shown that deviations from classical gravity in the bulk due to 
Weyl corrections in the Einstein-Maxwell action lead to a non-universal result for $k'/k$.  

\acknowledgments
The author would like to thank S. Sachdev for discussions and the 
Deutsche Forschungsgemeinschaft (DFG), grant No. KL 256/46-1, for 
the financial support.


\begin{thebibliography}{100}
\bibitem{Note-1} We have chosen here to follow Ref. \cite{Kovtun-Ritz} and include a factor $S_d^2$ in the definition of $k$ 
via the current correlation function.
\bibitem{Note-2} For the conformally invariant $O(n)$ model the current is given 
by $j_\mu^{ab}=\phi^a\partial_\mu\phi^b-\phi^b\partial_\mu\phi^a$ and 
$\langle j_\mu^{rs}(x)j_\nu^{tu}(0)\rangle=kS_d^{-2}x^{2(1-d)}
(\delta^{rt}\delta^{su}-\delta^{ru}\delta^{st})$, with the latin indices running from 
$1$ to $n$.
\end{thebibliography}
\end{document}